\documentclass[footinbib,aps,pra,reprint,superscriptaddress]{revtex4-1}
\usepackage{graphicx,tabularx} 
\usepackage[utf8]{inputenc}
\usepackage[T1]{fontenc}
\usepackage{float}
\makeatletter
\let\newfloat\newfloat@ltx
\makeatother
\usepackage{algpseudocode}
\usepackage{algorithm}
\usepackage{amsmath,amssymb,amsfonts}
\usepackage[colorlinks=true,bookmarks=false,citecolor=blue,urlcolor=blue]{hyperref}

\begin{document}

\title{Authentication of Smart Grid Communications using Quantum Key Distribution}

\author{Muneer~Alshowkan}
\email{alshowkanm@ornl.gov}
\affiliation{Quantum Information Science Section, Oak Ridge National Laboratory, Oak Ridge, Tennessee 37831, USA}
\author{Philip~G.~Evans}
\affiliation{Quantum Information Science Section, Oak Ridge National Laboratory, Oak Ridge, Tennessee 37831, USA}
\author{Michael~Starke}
\affiliation{Energy Systems Integration Section, Oak Ridge National Laboratory, Oak Ridge, TN 37831, USA}
\author{Duncan~Earl}
\affiliation{Qubitekk Inc., Vista, CA 92081, USA}
\author{Nicholas~A.~Peters}
\affiliation{Quantum Information Science Section, Oak Ridge National Laboratory, Oak Ridge, Tennessee 37831, USA}

\begin{abstract}
Smart grid solutions enable utilities and customers to better monitor and control energy use via information and communications technology. Information technology is intended to improve the future electric grid's reliability, efficiency, and sustainability by implementing advanced monitoring and control systems. However, leveraging modern communications systems also makes the grid vulnerable to cyberattacks. Here we report the first use of quantum key distribution (QKD) keys in the authentication of smart grid communications. In particular, we make such demonstration on a deployed electric utility fiber network. The developed method was prototyped in a software package to manage and utilize cryptographic keys to authenticate machine-to-machine communications used for supervisory control and data acquisition (SCADA). This demonstration showcases the feasibility of using QKD to improve the security of critical infrastructure, including future distributed energy resources (DERs), such as energy storage.
\end{abstract}

\maketitle

\section*{Introduction}

The electric grid is evolving from an electrical network composed primarily of large centralized fossil fuel plants to a more distributed infrastructure, which includes renewable and energy storage type plants. Wind, photovoltaic (PV), and energy storage system (ES) technologies have observed significant cost reductions as they have continued to mature and reach mass production~\cite{Yao2019,Cole2016,Stecca2020}. These technologies are now being adopted more frequently into the emerging electric smart grid, both in large and small deployments.

\begin{figure*}[tb!]
\centering
\includegraphics[width=\textwidth]{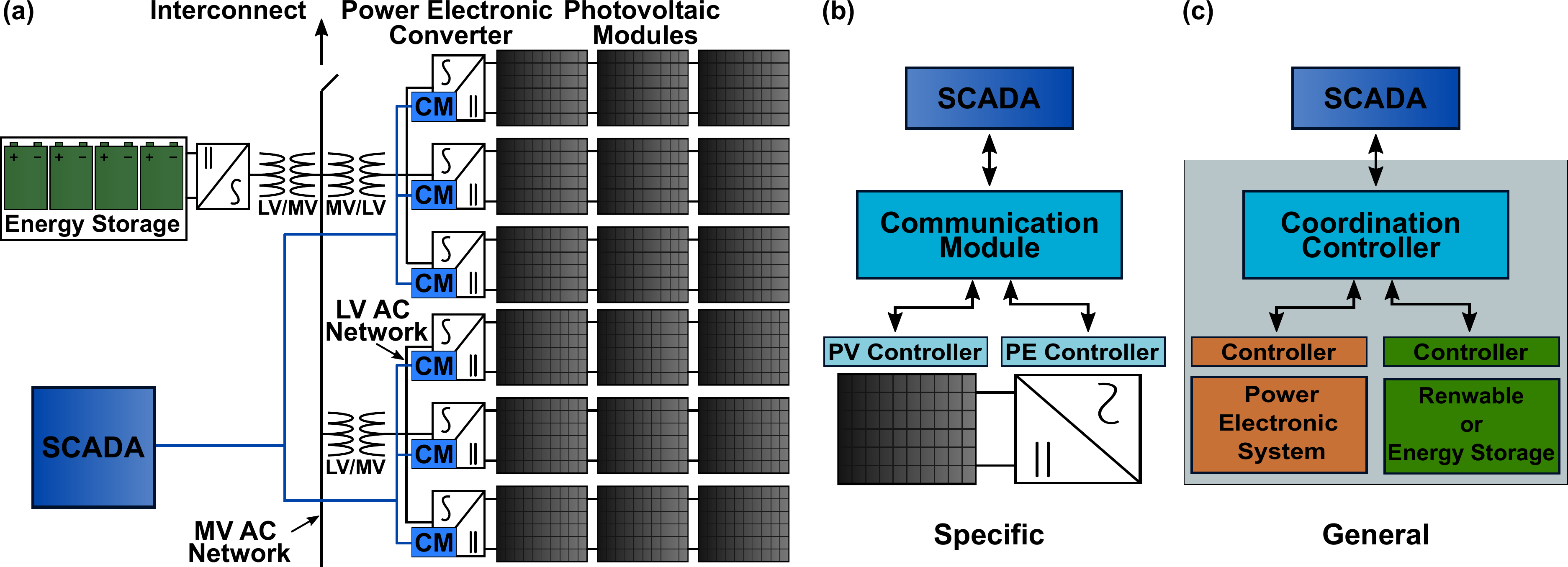}
\caption{(a)~Example of photovoltaic plant construction with voltage collector system (black) and communications network (blue). Architecture concept for (b)~Specific and (c)~General communications and control.
CM: Communications module.
LV: Low voltage. 
MV: Medium voltage.
PE: Power electronic.
PV: Photovoltaic.
SCADA: Supervisory control and data acquisition.
}
\label{fig01}
\end{figure*}

Renewable power plant installations can now be found on the scale of hundreds of kilowatts(kW) to megawatts (MWs) of potential power generation. These generation plants are a composite of many small generation resources, all interconnected with an electrical network known as a collector system~\cite{Camm2009a,Camm2009b,IEEE-2778-2020}. An example layout for a PV plant with a supplementary ES system is shown in Figure~\ref{fig01}(a). At each resource within the power plant, power electronic converter (PEC) systems with intelligent controllers are used to perform conversion and control of the power produced by both the PV modules and ES technology. These systems support several operational modes and communications protocols via an integrated communications module. System coordination is performed through a plant supervisory control and data acquisition (SCADA) system. Key to the deployment of these renewable plants is the ability for the SCADA system to communicate with the resources to establish operational capabilities and optimization strategies. Hence, secure and reliable two-way communications are critical to these systems~\cite{Tomsovic2005, Ma2013, Hossain2018}.

\begin{table*}[tb!]
\centering
\begin{tabular}{|c|c|c|c|c|c|c|c|c|}
\hline
\textbf{Protocol}       & \textbf{DNP3(-SA)}       & \textbf{EtherCat} & \textbf{FF HSE} & \textbf{IEC-60870} & \textbf{IEC-61850} & \textbf{Modbus} & \textbf{Powerlink}  & \textbf{Profinet}  \\ \hline
\textbf{Authentication} & No (DS) & No       & No     & DS        & DS        & No     & No         & DS       \\ \hline
\textbf{Open Source}    & Yes        & No       & Yes    & No        & YES       & Yes    & Yes        & No       \\ \hline
\textbf{Comm. Model}    & C/S        & C/S      & C/S    & C/S       & C/S       & C/S    & C/S \& P/C & P/C      \\ \hline
\textbf{TCP/UDP}        & both       & both     & both   & TCP       & TCP       & both   & both       & both      \\ \hline
\end{tabular}
\caption{SCADA communications protocols and their characteristics.
C/S: Client/Server communications model.
DS: Digital Signature.
DNP3: Distributed Network Protocol 3.
DNP3-SA: Distributed Network Protocol 3 Secure Authentication.
FF HSE: Foundation Fieldbus High Speed Ethernet.
IEC: International Electrotechnical Commission.
P/C: Producer/Consumer communication model.
SA: Secure Authentication.
}
\label{SCADAprotocols}
\end{table*}

Within a conventional SCADA system, a supervisory system, a human-machine interface (HMI), a communications network, a master terminal unit (MTU), remote terminal units (RTUs), and field devices. Hence, the communications network enables connectivity between the systems. Moreover, a SCADA communications network can be divided into four types: (1) monolithic systems that are isolated and do not interact with one another, (2) distributed systems that communicate over a local area network (LAN), (3) networked systems that operate in multiple sites and communicate over a wide area network (WAN), and (4) Internet of things (IoT) systems that are connected to cloud computing for widescale implementation and computational resource availability. Furthermore, the need for reliable, efficient, and continuous connectivity between the SCADA elements has led to the development of many different communications protocols. 
Some protocols have been designed to consider the processing power and communications requirements of industrial applications, while others focused on speed. Consequently, many protocols were designed without integrated security services such as authentication and encryption.
While the SCADA system in the monolithic and distributed models can operate in isolation on private links, utilities are looking to use available or existing communication infrastructure such as WANs and IoT to reduce costs which are often shared with other entities or service providers. Consequently, communications in these models are vulnerable to cyberattacks. For instance, the well-known ethernet-based SCADA communication protocols such as DNP3, EtherCat, Powerlink, Foundation Fieldbus HSE, and Modbus do not offer any authentication security mechanism. On the other hand, protocols such as DNS3-SA, IEC-60870, IEC-61850, and PROFINET implement security measures based on digital signatures. Table~\ref{SCADAprotocols} shows the characteristics of these protocols, and a comprehensive review of SCADA communication protocol and their security can be explored in~\cite{Pliatsios2020}.

In addition to these standard communication protocols, IoT protocols such as message queuing telemetry transport (MQTT), data distribution service (DDS), hypertext transfer protocol (HTTP), constrained application protocol (CoAP), and advanced message queuing protocol (AMQP) can be implemented in SCADA systems for machine-to-machine (M2M) communications. MQTT~\cite{Mqtt-v5} is a valuable protocol in the context of the IoT. MQTT has been utilized by companies such as IBM, Microsoft, and Amazon to operate as a message server that connects cloud applications and IoT devices. In comparison to SCADA systems, this protocol is similar to those often used in that data is frequently sought from other stations. One advantage of MQTT is that the protocol can be used with edge devices to integrate with older systems. Control stations and remote devices may be detached and communicate only over MQTT. Therefore, this simplifies peer-to-peer communications and relieves control stations of middleware duties. For this reason, MQTT has been recently explored and prototyped for SCADA systems~\cite{Jamborsalamati2019, Kodali2016, Starke2019a, Starke2019b, Starke2020, Starke2021a, Starke2021b}.

As presented in~\cite{Ghosh2019}, SCADA systems have been the target of many attacks that can impact the reliability of the communications network. These attacks include eavesdropping, man-in-the-middle, masquerade, virus and worms, trojan horses, and denial-of-service. These attacks have targeted the various levels of SCADA networks including the application layer, session layer, network transport layer, data link layer, and physical layers, with varying success rates. Therefore, electric utilities and generation plants are applying many different approaches to secure the information flow. These methods include adopting considerations of privacy/confidentiality, integrity, authentication, and trusted computing~\cite{Zhu2011,Yan2012,Ghosh2019}.

Solutions for ensuring the privacy and integrity of the communicated data include utilizing encryption and authentication. Both encryption and authentication schemes use cryptographic algorithms and secret keys. However, the two general schemes are different: encryption converts a message plaintext to a ciphertext to protect the information, whereas authentication is the attribute of confirming a message is genuine and has not been altered during transmission.

Currently, many popular cryptographic solutions, such as public-key cryptography, are based on hard-to-solve mathematics using assumptions based on potentially available computing resources~\cite{Shor1997, Scarani2009}. One of the major advantages of public-key cryptography is enabling messages to be encrypted and/or authenticated with a ``public'' key (i.e., known to all) which in turn can only be decrypted and/or signed with a ``private'' key (i.e., kept secret). The generation of the public-private key pair leverages the aforementioned mathematics. To continually improve security of this type of cryptography, the secret key size must increase with available computational capabilities~\cite{Barker2015}. This can be a challenge for devices deployed in the field as the availability of computational resources (i.e., memory size and processing capability) is typically fixed during deployment or when the device is constructed. Hence, without detrimentally increasing latency or potentially being put out of service---as the processing demand increases---devices in the field must be replaced~\cite{Hauser2012, Wei2011}.

In contrast, private-key cryptography---where a single key performs both encryption and decryption tasks---can be implemented very efficiently in hardware~\cite{Mangard2003}, while exhibiting low computational overhead with deterministic latency. However, the challenge is all keys must be securely distributed to all parties prior to use, typically by a trusted courier, resulting in all keys being at risk of discovery during transit. From this perspective, quantum key distribution (QKD) approaches offer considerable promise: keys for private-key cryptography schemes can be established between parties---even over communication channels controlled by an adversary---in a provably secure manner~\cite{Lo1999}. Arguably, QKD is one of the most mature quantum applications available~\cite{Scarani2009}. The fundamental technology has already been observed to be transitioning from research laboratories to commercial products. Combined with information-theoretic security protocols~\cite{Shannon1948}, QKD offers future-proof security: proven to be safe regardless of the technological development in computing, quantum or otherwise~\cite{Scarani2009}.

Quantum Key Distribution describes a variety of techniques whereby quantum states are used to establish a shared random key between two spatially separated parties, commonly referred to as \textit{Alice} and \textit{Bob} in cryptographic parlance. BB84~\cite{Bennett2014} is the most well-known QKD protocol, yet others exist which leverage different encoding schemes~\cite{Bennett1992,Scarani2004} as well as entanglement~\cite{Ekert1991}. QKD is not a cryptographic mechanism---it is a method to distribute correlated random bit strings for later use in any application, including well-known symmetric cryptography schemes such as the Advanced Encryption Standard (AES), Blowfish, and others. The commercial QKD system used in this paper implements an entanglement-based protocol~\cite{Ekert1991}. It generates keys that are pulled into a higher layer to authenticate smart grid communications.

Securing a simulated power grid communications network using QKD was presented in~\cite{hughes2013} and using real time digital simulator (RTDS) microgrid testbed in~\cite{Tang2021} while theoretical approaches to improve the power grid physical security using quantum computing were explored in~\cite{Eskandarpour2020}. Previously, QKD has been applied in a trusted relay testbeds~\cite{Elliott2002, Peev2009, Chen2010, Stucki2011, Sasaki2011, Dynes2019, Chen2021} as well as a fiber loop-back on a utility network~\cite{Evans2019}. Following the initial utility demonstration, a four-node QKD trusted relay network on a utility fiber infrastructure showed the interoperability between diverse QKD systems that worked together to deliver secure keys across the critical energy infrastructure~\cite{Evans2021} using the one-time-pad encryption technique. In~\cite{Chen2021} the secret keys were further used to encrypt banking communication systems via the AES-128 protocol. Hence, authentication---which is a fundamental cryptographic security service---of typical network communications was not demonstrated in any previous work to secure the power grid communications as the secret keys in the trusted relay experiments were used only for encryption of distributed keys to relay them between the network nodes.

Our main objective is to achieve in principle information-theoretic authentication in smart grid communications. Our specific implementation uses the publish-subscribe paradigm, which is popular for smart grid data, and in particular the MQTT protocol. We develop a detailed methodology, practical design, and integrate several heterogeneous components on each publisher-subscriber link in the deployed energy delivery infrastructure. The major challenges to realizing authentication are the commodity SCADA microcontrollers' limited resources, as well as their integration with a QKD system and the quantum random number generators (QRNG). Additionally, a further challenge we solve is how to manage the random numbers and the secret keys over the distributed devices.

While a review of the challenges of using QKD in the context of smart grid communications has been explored in~\cite{Peng-Yong2022}, here we highlight the challenges related to securing the SCADA communications and the concepts developed to accomplish this task in our demonstration. 
One challenge with using public networks like WANs in the smart grid is that the networking infrastructure is often shared. 
A challenge arises when data leaves the utility network and becomes vulnerable to cyberattacks. 
A network design must be developed to provide authentication and verification services to real-time outgoing and incoming communications messages. The lack of integrated security services---such as authentication and encryption---is another challenge associated with many existing SCADA communications protocols. 
As a result, these protocols are also susceptible to cyberattacks.
Although some protocols rely on computationally intensive public-key digital signatures for authentication, the length of their secret keys must be increased to maintain their security over time. 
Devices in the field often face this challenge because the computational resources available after deployment are often fixed.
Moreover, SCADA systems utilize specialized microcontrollers with limited resources that may be incapable of performing the intensive calculations required for public-key cryptography as key sizes increases. Therefore, equipment in the field must be upgraded to prevent communications delays and outages. This is a challenge for devices that are deployed in remote locations and are intended to operate for a long time.

To overcome these challenges, we present specialized and generalized architectures in which QKD secret keys protect SCADA communications. The generalized approach can be applied for proprietary protocols, including many-to-many communications scenarios. The specialized network architecture intends to operate effectively for open-source point-to-point communication protocols. Utilizing the open-source MQTT protocol---which can be used for an edge device and can be integrated with older systems---is a concept that provides flexibility in terms of communications and security. Consequently, a compatible, lightweight, and information-theoretic authentication protocol can be incorporated into MQTT and operated on the SCADA microcontrollers, reliably performing authentication and verification services. Furthermore, we solve the latency challenges with private-key cryptography, in which a single key performs encryption and decryption functions with minimal computing overhead and delays. Using quantum key distribution (QKD) techniques, secure keys for private-key cryptography schemes can be established between participants. We integrate QKD keys in information-theoretically secure protocols to provide a future-proof authentication that is secure and independent of the advancement of classical or quantum computing technology. Therefore, our computationally efficient approach is able to overcome the challenges associated with limited computing resources as the key size increases in public-key cryptography. We compare the execution time of our technique to the public-key cryptography counterpart, demonstrating its feasibility for smart grid applications and showing how QKD can benefit grid communications.

In this paper, we achieve our objective by using QKD secret keys to authenticate communications of integrated power electronics energy resources in electric grid infrastructure. This work is the first time quantum secret keys have been used to authenticate smart grid communications. More specifically, (a) QKD secret keys have been applied over the IoT protocol MQTT for supporting DER communications, (b) the developed software design to utilize and manages secret keys established by a commercial Qubitekk quantum key distribution system to authenticate M2M communications, and (c) the platform has been applied in a real utility setting (at EPB in Chattanooga Tennessee, between a data center and an electrical substation connected via an optical fiber). We first lay the foundation of our developed approach in the next section and then provide a detailed description of our system and methods used to solve the challenges in the following sections.

\subsection*{Message encryption and authentication in QKD - Galois message authentication }
\label{GMAC}
The concept of provably secure authentication was introduced in~\cite{Gilbert1974} using a secret key that is longer than the message itself. Carter and Wegman showed it is possible to use a secret key shorter than the message to achieve information-theoretic authentication~\cite{Wegman1979}. Later, using a block cipher, it was shown by Brassard that a shorter secret key could be expanded and used for the Carter-Wegman authentication scheme~\cite{Brassard1983}. Galois/Counter Mode (GCM) is a state-of-the-art parallelizable symmetric-key cryptographic protocol based on the Carter-Wegman authentication scheme~\cite{mcgrew2004}; it offers information-theoretic encryption and authentication. The Galois Message Authentication Code (GMAC) is the GCM standalone authentication scheme, i.e., where the message does not need to be encrypted. The National Institute of Standards and Technology (NIST) approved GCM and GMAC in 2007 via NIST SP 800-38D standard~\cite{dworkin2007sp} which is also part of the federal information processing standards (FIPS).

There are three inputs to the GMAC: (1) the message to be authenticated, (2) an initialization vector (IV), also referred to as a \textit{nonce}, and (3) a secret key. The output is the message authentication code (MAC). As expected in symmetric-key algorithms, GMAC assumes a fundamentally secure key exchange between the sender and the receiver. GMAC allows reusing a secret key to authenticate more than one message; however, it prohibits using it with the same nonce~\cite{dworkin2007sp}. Currently, the acceptable block ciphers recommended by NIST are AES-128, AES-192, and AES-256~\cite{barker2018}. For the nonce, the acceptable size is 96 and 128-bits. The length of the output message authentication code is 128 bits. The authentication process is initiated by a sender (Alice) who wants to send an authenticated message to a receiver (Bob). A new secret key, a nonce, and the original message are then supplied to the GMAC, which outputs the message authentication code. Alice sends the original message, the nonce, and the MAC to Bob but keeps the secret key a secret. Upon receipt, Bob then forwards Alice's message, nonce, and MAC along with the corresponding secret key to the GCM verification algorithm, whose output is a simple statement: true if the message is authentic or false if not.

\begin{table*}[th]
\setlength{\tabcolsep}{9pt}
\renewcommand{\arraystretch}{1.3}
\centering
\begin{tabular}{|c|c|c|}
\hline
\textbf{Publisher} & \textbf{Subscriber} & \textbf{Data} \\ 
\hline
SCADA & \begin{tabular}[c]{@{}c@{}}Coordination \\ Controller\end{tabular} & \begin{tabular}[c]{@{}c@{}}Topic: PV/Control \\ Payload: Control Information\end{tabular} \\ 
\hline
\begin{tabular}[c]{@{}c@{}}Coordination \\ Controller\end{tabular} & SCADA & \begin{tabular}[c]{@{}c@{}}Topic: PV/Measurement \\ Payload: Measurement Information\end{tabular} \\ 
\hline
\end{tabular}
\caption{SCADA and DER communications using publisher-subscribe model}
\label{table01}
\end{table*}

\section*{Communications and control architecture}
In this work, the concept of operations is the communications between a single photovoltaic (PV) system and a SCADA system. In the following sections, a generalized architecture for supporting the authentication of smart grid communications using quantum key distribution demonstration is discussed. 

The integration of a power electronic controller (PEC) and an energy resource to construct a distributed energy resource (DER) can be performed through a multiple vendor ``black-box'' integration effort~\cite{Arnedo2009, Valdivia2009}. The ``black-box'' designation signifies that only a communications interface to the system is present, as shown in Figure~\ref{fig01}(b). This work proposes an architecture that utilizes an integration layer (or coordination controller) to couple systems and providers, shown in Figure~\ref{fig01}(c). The proposed coordination controller can be placed directly within the hardware system and provides an opportunity to automatically enable QKD systems to be applied to many different PEC-type resources.

The coordination controller has been developed as a means to integrate many types of PECs and resources. The design utilizes a multi-agent architecture comprised of four agents: converter, source/load, interface, and intelligence. The \textbf{Converter Agent} interacts with the PEC then shares the status and data over a local messaging bus. The \textbf{Source/load Agent} interacts with the source/load then transmits data which includes control and status, with other agents. The \textbf{Interface Agent} interacts with the external agents to send and receive information, then relays the information to the local agents over the local message bus. Finally, the \textbf{Intelligence Agent} interacts with the interface agent to convert requested control signals into actionable signals for the separate resources. All communications between agents and message buses utilize the MQTT protocol. As an example, a start-up request is broken into manageable steps between the resource and the PEC to complete the task. These operations must be tightly synchronized and often autonomous to avoid errors and protect the energy infrastructure. This approach has been demonstrated in the development of energy storage systems and PV from residential~\cite{Starke2019a, Starke2019b, Starke2020, Starke2021a}, to commercial-scale~\cite{Starke2021b} systems in both hardware and controller hardware-in-the-loop platforms. We note that other MQTT-based work for autonomous resource allocation systems was explored in~\cite{Jamborsalamati2019} and for automation systems in~\cite{Kodali2016}.

In this work, an MQTT messaging approach between the SCADA system and the DER coordination controller is outlined as follows: the SCADA subscribes to measurement data published by the DER coordination controller and the DER coordination controller subscribes to control data published by SCADA as presented in Table~\ref{table01}. An example of an auto-commissioning sequence through a registration process is presented in~\cite{Starke2019b}.

\begin{figure*}[t!]
\centering
\includegraphics[width=\textwidth]{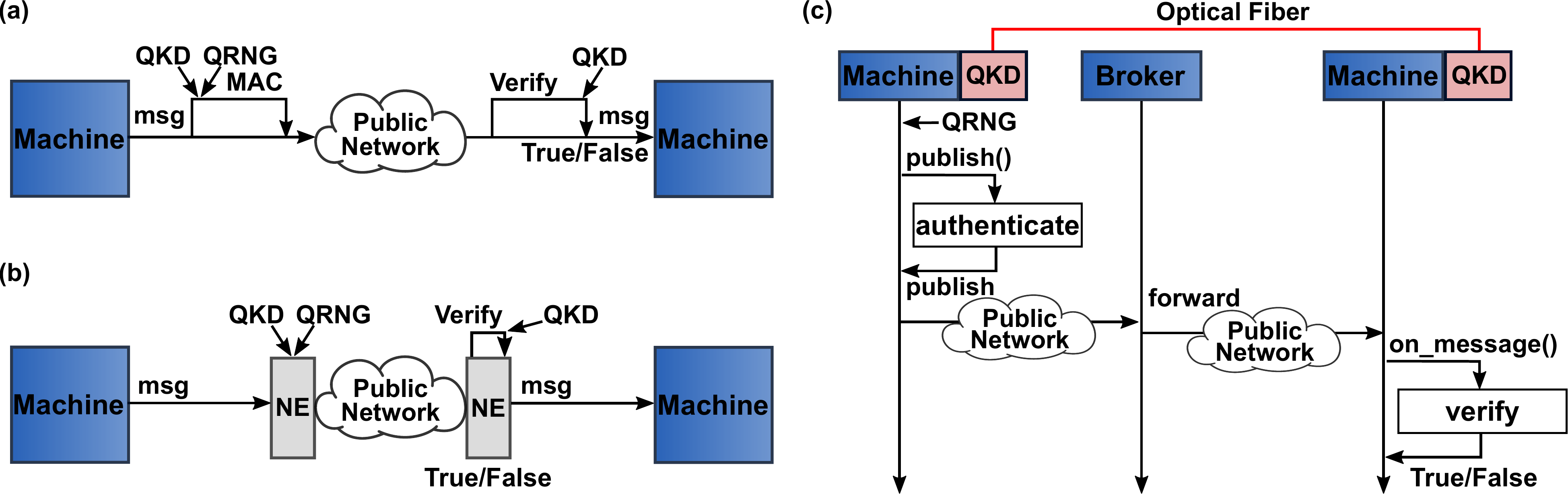}
\caption{Showing (one-way for simplicity) generalized authentication approach using QKD secret keys and QRNG that can be implemented in: (a)~Open-source and (b)~Proprietary SCADA communication protocols to authenticate outgoing messages by one machine and verify incoming messages by another. (c)~Implementation of the authentication approach using QKD secret keys and QRNG in the MQTT publisher/subscriber protocol operating for SCADA communications.
MAC: Message authentication code.
msg: SCADA message. 
NE: Network encryption card.
on\_message(): MQTT specific on received message callback function.
publish(): MQTT specific on publish message callback function.
QKD: Quantum key distribution.
QRNG: Quantum random number generator.
}
\label{General-QKD}
\end{figure*}
\vspace{-0.01in}

\section*{Methods: Applied QKD Key Authentication Approach}
This section describes the authentication methods integrated into the MQTT-based protocol M2M SCADA communications system. We show the applied QKD keying approach to achieve information-theoretic authentication communications between publisher and subscriber.

\subsection*{Generalized Authentication Approach}
In general, machine-to-machine authentication can be accomplished by creating a cryptographic challenge using secret keys that are only known to the sender and the receiver. Ideally, through information-theoretic concepts combined with secret keys distributed over communication networks via QKD. Assuming the SCADA communication protocol is open-source, it is possible to implement such an authentication protocol for each outgoing message by sending the original message accompanied by its challenge (e.g., MAC). Then, the receiver uses a verification function to check the authenticity of each received message. This verification function will enable the SCADA receiving machine to accept or reject the received message, as shown in Figure~\ref{General-QKD}(a). For proprietary SCADA communication protocols, QKD secret keys can be used in network encryptors modules as shown in Figure~\ref{General-QKD}(b) that perform end-to-end encryption and authentication services~\cite{Choi2014,Dynes2019,alshowkan2021advanced}. A Benefit of this approach is solving challenges in the system scalability, as described in~\cite{Peng-Yong2022}. In this case, the traditional point-to-point QKD system---including long-distance deployment via satellite---can be facilitated for many-to-many communications models.

\subsection*{Specific Authentication Approach to MQTT}
Assuming Alice and Bob share a set of QKD-based secret keys $k_1,...,k_n $ where $n$ is an arbitrary serial number for each key. To guarantee that only one user uses each secret key, we give each secret key a serial number. Then we assign secret keys with odd serial numbers $k_{odd}$ to Alice and secret keys with even serial numbers $k_{even}$ to Bob. Moreover, we also assume that each user has a set of random initialization vectors $ iv_1,...,iv_j $ privately generated from a quantum random number generator where $ j $ is an arbitrary serial number for each $ iv $. 
To publish an authenticated message $m$ and its topic $t$ ---which is an MQTT specific variable and part of every packet---using a secret key $k_n$, the secret key serial number $n$ used in this process need to be transmitted to indicate to the receiver which key was used (without disclosing any information about the key itself).
In our case, we choose to set the key serial number to be part of the overall message to be authenticated.
To avoid replay-attacks, an authenticated timestamp $ts$ is used. Therefore, the total message $tm_i$ where $ i $ is the number of the message and it's related topic to be authenticated becomes:

\begin{equation}
    tm_i = m_i + t_i + n +ts
\end{equation}

\noindent
In our software, we utilized the MQTT built-in \emph{publish()} callback function as shown in Figure~\ref{General-QKD}(c) to create the specific message authentication code from the sender $mac_S$ for each $tm_i$ being published using the GMAC encryption $GMAC_E$ algorithm such that: 

\begin{equation}
    mac_S=GMAC_E(tm_i,k_n,iv_j)
\end{equation}

\noindent
where the total message $tm_i$, secret key $k_n$, and initialization vector $iv_j$ are inputs to the GMAC algorithm, and $mac_S$ is a 16-byte string uniquely associated with the inputs. Once $k_n$ and $iv_j$ are retrieved for use with the GMAC, they are immediately flagged as used. To verify the authenticity of the $tm_i$, Alice needs to share the $mac_S$ and the initialization vector $iv_j$ with Bob while keeping the secret key $k_n$ secret. Thus, the payload $p$ of each message being published becomes

\begin{equation}
    p=tm_i+iv_j+mac_S
\end{equation}

While the payload of a standard MQTT message contains only the message data $ m $, we employed a delimiting character between the components of the total message $tm_i$ to construct the new payload (e.g., using dashes, $ m_i-t_i-n-ts-iv_j-mac_S$) for convenient payload coding and decoding. We utilized the MQTT \emph{on\_message()} callback function to verify every received message. For each received message, we use the delimiting character to break the payload data---to retrieve all the components of the total message $tm_i$---and start the verification process. First, we verify using $k_n$ that the secret key has not been used before. Second, comparing the last used $k_n$ and the $ts$, we verify that the message is not delayed or replayed by considering the typically expected delays in the network $\delta$. While the $ts$ will depend on classical network synchronization (e.g., Precision Time Protocol and Network Time Protocol), any anomalies detected in timing between the nodes will trigger further investigation. Third, we use the message topic and verify it is equal to the topic embedded in the $tm_i$. Fourth, using the received $tm_i$, $iv_j$, $mac_S$, and the corresponding $k_n$, the receiver performs the verification GMAC decryption $GMAC_D$ as follows:

\begin{equation}
    mac_R=GMAC_D(tm_i,k_n,iv_j)
\end{equation}

\noindent
Bob compares the received 16-byte $mac_S$ and the calculated $mac_R$. If both match, then $tm_i$ and subsequently the original message $m_i$ are authentic, otherwise the authenticity cannot be established for this message, and further investigation is warranted. Upon successful verification, Bob flags the received key as used. Supplementary Algorithm~1 and Algorithm~2 summarize the process of creating and verifying the MAC, respectively.

\begin{figure*}[!t]
\centering
\includegraphics[width=\textwidth]{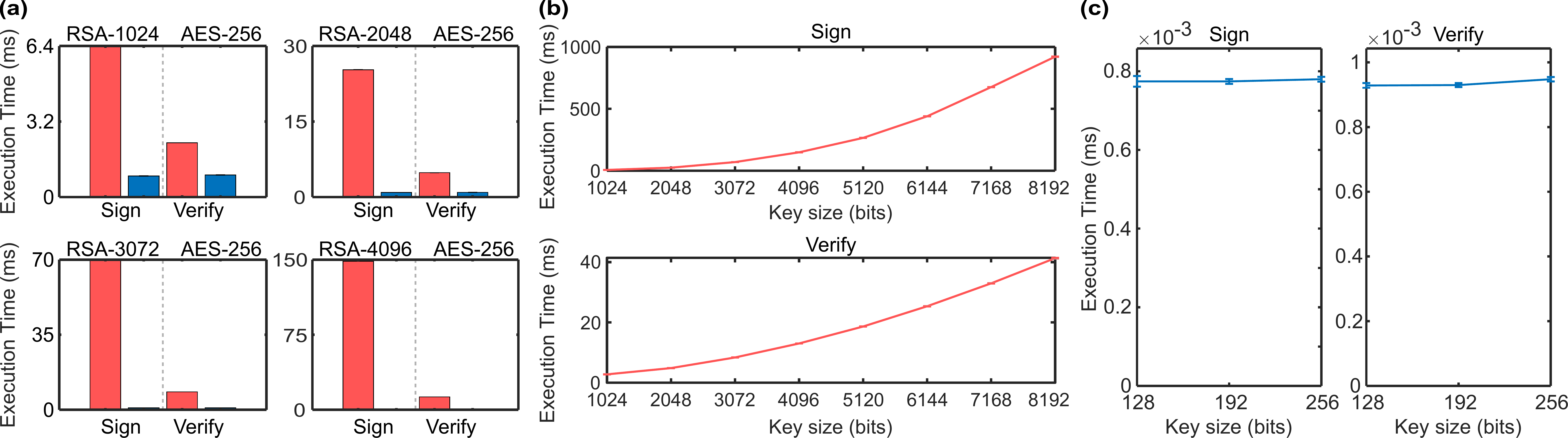}
\caption{Measurement results for GMAC (blue) and Digital Signature (red) for (a)~Execution times to authenticate a 256-byte message using GMAC and digital signature based on RSA with 1024, 2048, 3072, and 4096-bit and AES-256 bit keys. The vertical dotted lines are for visualization only to separate signing from verifying bars. 
(b)~An extended measurement to show the trend of execution time for larger key sizes for RSA-based digital signature. 
(c)~Authentication computation time of a 256-byte message using AES-based GMAC with s key sizes of 128, 192, and 256-bits.
}
\label{ex_time}
\end{figure*}

\subsection*{Device Execution Time Measurement}
Theoretical information on the complexity of the underlying cryptographic algorithms has already been explored and can be found in~\cite{boneh1999twenty,dworkin2007sp,bogdanov2011biclique}. Therefore, in this section, we characterize the device running the Python-based DER system by measuring the authentication execution time in the same programming language. Each DER machine runs on Raspberry Pi 3b+, which is equipped with a 1.4GHz Cortex-A53 quad-core processor and 1GB LPDDR2 SDRAM. We compare the proposed authentication using GMAC to the digital signature available in some SCADA communications protocols. Because SCADA messages are typically short, we set the message length to 256 bytes in the following measurements.

Figure~\ref{ex_time}(a) shows the execution time to sign and verify a message using digital signatures based on RSA 1024, 2048, 3072, and 4096-bit keys compared to GMAC based on AES with a 256-bit key and 128-bit nonce---the longest recommended secret key and nonce by NIST~\cite{barker2018}. The average execution times in milliseconds (ms) to sign (verify) a message using 
\textbf{GMAC with AES-256:} $0.8895 \pm 0.0072$ ($0.9309 \pm 0.0088$),
\textbf{RSA-1024:} $6.3507 \pm 0.0137$ ($2.2864 \pm 0.0037$), 
\textbf{RSA-2048:} $25.2802 \pm 0.0214$ ($4.8489 \pm 0.0057$),
\textbf{RSA-3072:} $69.9515 \pm 0.0450$ ($8.3635 \pm 0.0071$) and, 
\textbf{RSA-4096:} $148.4858 \pm 0.0207$ ($12.9215 \pm 0.0078$).
The uncertainties are reported as the standard deviation of the mean of 512 samples. While the currently recommended RSA key sizes are 2048 and 3072-bit, we are showing the measurement results for RSA 1024 and 4096-bit to illustrate the execution time of the previous and possibly future RSA standards~\cite{Barker2015}, respectively. We notice that, as the RSA key size increases, the execution time significantly increases and note that at a delay of 160~ms (total time to sign and verify a message for the RSA 4096) it is possible to get electric grid synchronization errors. An extended measurement to show the trend of execution time for larger key sizes for RSA up to 8192-bit is shown in Figure~\ref{ex_time}(b).
Figure~\ref{ex_time}(c) shows the GMAC authentication execution time of a message using AES with key sizes of 128, 192, and 256-bit (the maximum possible key size to measure), with a maximum delay of less than 2~ms to sign and verify. In contrast with the RSA results, the GMAC results show a negligible increase in the signing and verification times, indicating that increasing the key size in the future is feasible with negligible added delay.

\begin{figure*}[t!]
	\centering
	\includegraphics[width=\textwidth]{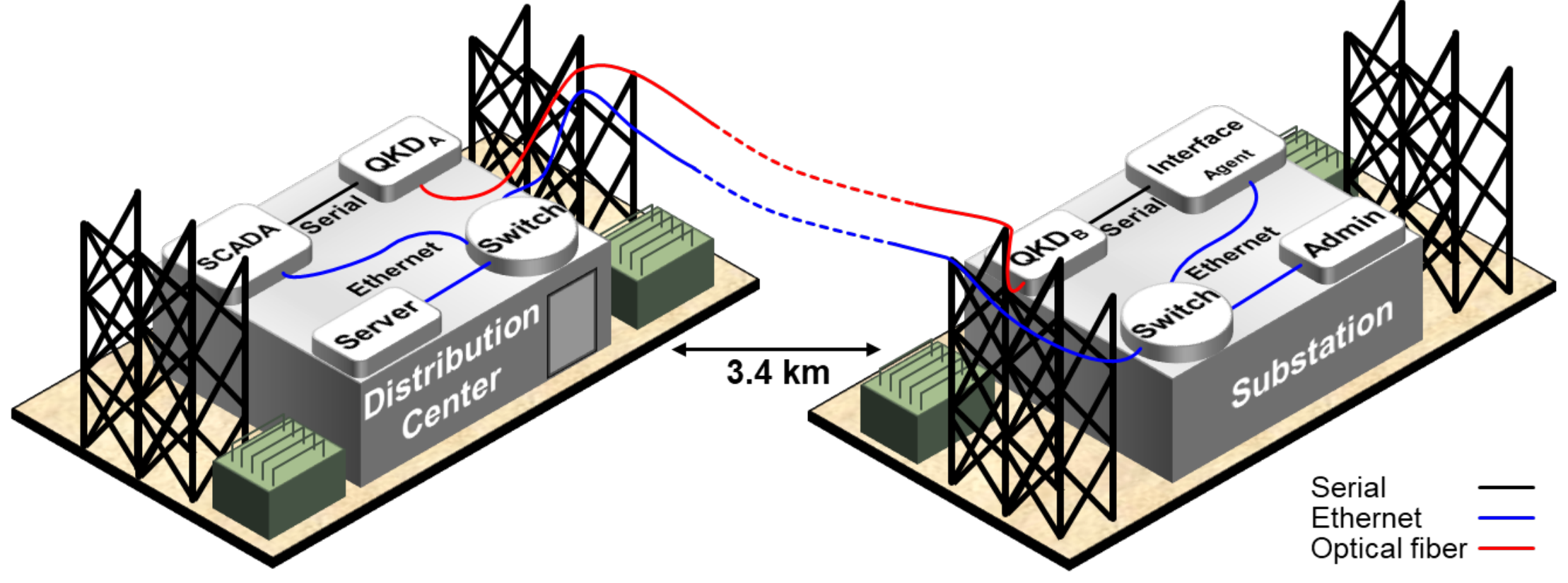}
	\caption{Illustration of the network configuration. The distribution center contains the QKD Alice system, SCADA (Supervisory control and data acquisition) system,	and a server to collect network statistics. The substation contains the QKD Bob system, the energy storage system Interface agent, and an admin computer for real-time monitoring.}
	\label{fig04}
\end{figure*}

\subsection*{Demonstration on the electric grid}
We demonstrate the above QKD-enabled MQTT approach in a real-world electrical utility environment at the Electric Power Board (EPB), Chattanooga, Tennessee. Two optical fibers are used to create a dedicated quantum communications link between a distribution center (DC) and an electrical substation (SUB). A Qubitekk Industrial Control Systems (ICS) commercial QKD system is used in this demonstration. The link distance between DC and SUB is approximately 3.4 km and exhibits an optical attenuation of 1.3 dB at 1550 nm, including patch panel connectors and splices. While the dedicated optical fiber is bundled with many other optical fibers used for utility operations, we note that the quantum communications link and all other classical communications links used for this work are isolated from EPB's operational network. This isolation is good practice for testing experimental technologies in operating power grid infrastructure. In this network, the bulk of the optical fiber link is deployed aerially between utility poles and hence experiences environmental variables such as temperature changes and wind motion. This in turn, has a slight effect on the quantum key generation rates, as would be expected with polarization encoded photons utilized by the Qubitekk system. In addition to the dedicated quantum optical fiber links, we also establish a typical TCP/IP local area network for the corresponding classical channels between virtual distributed energy storage systems located at DC and SUB.

\subsection*{Network Configuration}
The QKD hardware is deployed at the utility between DC and SUB. At each location, a virtual distributed energy resource (vDER) machine on Raspberry Pi 3 B+ was deployed: the Intelligence (Intel) Agent machine is set up in DC, and the Photovoltaic (PV) Agent machine in SUB. Each system is connected to the private classical network via a network switch (see Figure~\ref{fig04}). Additionally, two other supporting devices in the same network are connected: 1) a server to collect the network statistics in DC and 2) a device in the substation, used for administration tasks, including control and data monitoring.

\subsection*{Authentication Software}
\label{demo_software}
Software that handles network node secret key and random number operations, including retrieving, verifying, and managing the materials, has been developed for this demonstration. Then, we utilize these materials to authenticate the vDER communications. While performing these operations, each node is responsible for tracking and reporting statistics related to the secret keys, random numbers, and the completed tasks. Because the communications in this network follow a publish-subscribe architecture; when the software starts, the receiving node verifies the authenticity of the transmitting node; then subscribes to topics of interest. In what follows, the basic functionalities of the network nodes are described.

\begin{figure*}[!t] 
\centering
\includegraphics[width=4.5in]{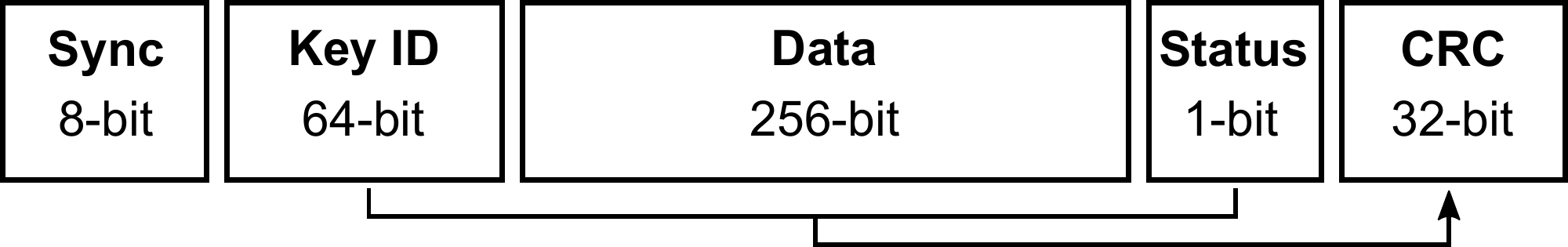}
\caption{The format of the secret keys retrieved from the QKD system. Sync: Synchronization, a 8-bit to determine the start of a frame. Key ID: 64-bit key counter. Key Data: 256-bit Secret key. Key Status: 1-bit status of the key. CRC: Cyclic Redundancy Check, a 32-bit for key id, key data, and key status.}
\label{figt}
\end{figure*}

\subsection*{Secret key management}
\label{key_management}
Python-based software runs a background service in each device, retrieving secret keys from each QKD system over a serial cable. As the keys become available, the background software stores them in a local file. Figure~\ref{figt} shows the format of the key materials retrieved from the QKD system and stored in the local file. 
One functionality of the software that has been developed and integrated into the vDER system is the periodic monitoring and data retrieval of new key materials from the local file as they become available. The software checks every received key for the appropriate length to avoid run-time errors caused by inadequate key length. In this case, the key of length 32 bytes (256-bit) is verified as a valid key. Finally, the software creates a record for each key, including a serial identification number and a Boolean status flag indicating the used and unused secret keys. After this point, each node should have an identical key table to use for MQTT protocol communications authentication.

\subsection*{Random number management}
\label{qrng_management}
Like secret key management, each node has access to a list of local (initially private) random numbers generated from a quantum random number generator to use as initialization vectors. In our case, we use random numbers generated from a commercial IDQ QRNG. The QRNG outputs a large string of random numbers that we chunk into smaller strings, each of length 16 bytes (128-bit) which the authentication algorithm accepts. Because these random numbers do not need to be identical between the network nodes, each node manages them locally. When a node plans to create a new MAC for a message, a random number from the local list is retrieved and a corresponding flag is set to ``used'' to never be used again.

\subsection*{Authentication and verification}
Authentication and verification are the core of the software integrated in MQTT-based SCADA system described in the previous section.
The software is called when the vDER systems want to publish a message to create its MAC. The original message gets appended with the MAC and other supporting information to enable a receiver that shares secret keys with the sender to verify the message's authenticity. Additionally, the software asserts other security measures to prevent replay and delay attacks. For this reason, the timestamp, message topic, and secret key serial identification number are also set to be authenticated and verified by the receiver. Thus, a received message gets verified against replay and delay attacks. For example, the software verifies a timely message receipt by tracking the last secret key used, confirming the expected behavior of message sequence, in addition to checking the timestamps.

\subsection*{Statistics reporting}
For monitoring, each node periodically reports general information to the statistics server. For example, information reported related to the key management includes the number of available, added, and used secret keys. Similarly, information related to the random numbers, including those added, available, and used, is also reported. Additionally, the verification algorithm reports the number of successful and unsuccessful message verification instances.

\section*{Results}
Using the developed authentication approach in the MQTT protocol operating in the SCADA system and the software described above, we authenticate the communications between the PV and Intel agents using secret keys from the deployed QKD system. When the PV and Intel agents start, they proceed to perform the secret key and random number management described in the previous section.
A set of global variable objects are initialized of various classes needed to support the communications, interfaces, and measurements. Additionally, if enabled, the agents set the graphical user interface (GUI) parameters. After the initialization step, each agent requests to connect to the broker using the broker IP address and port number---the default MQTT port number is 1883---over the TCP/IP protocol. A successful connect request by an agent is acknowledged with a message containing a connect flag by the broker. Individually, each agent informs the broker about the list of topics of interest and each quality of service (QoS). The QoS indicates the level of reliability required based on the network and the application requirements. QoS 0 indicates a best-effort service---delivery is not guaranteed---a published message is transmitted to a subscriber once, and no acknowledgment of delivery is required. In QoS 1, a published message is generated to be delivered at least once. Therefore, an acknowledgment flag is required from the subscriber to confirm the delivery, or retransmission of the same message is triggered: lost acknowledgment flags trigger retransmission of previously delivered messages. Using a four-way handshake, QoS 2 guarantees that a message is published and delivered to a subscriber exactly once: it ensures that no duplicate messages are sent to the same client. Hereafter, agents in the vDER software are connected to the broker and subscribed to each other's topics. Consequently, their published messages are authenticated and verified using the QKD secret keys. 

The published messages between the agents include slow and fast local periodic messages. For example, the Intel agent publishes control and request information related to the type of the system on the identification of need, such as control and setpoints, while the PV agent publishes configuration and forecast using a slow periodic timing (still in seconds). On the other hand, the PV agent publishes the system status, measurements, and errors in the fast periodic local messages.

\begin{figure*}[!t]
\centering
\includegraphics[width=\textwidth]{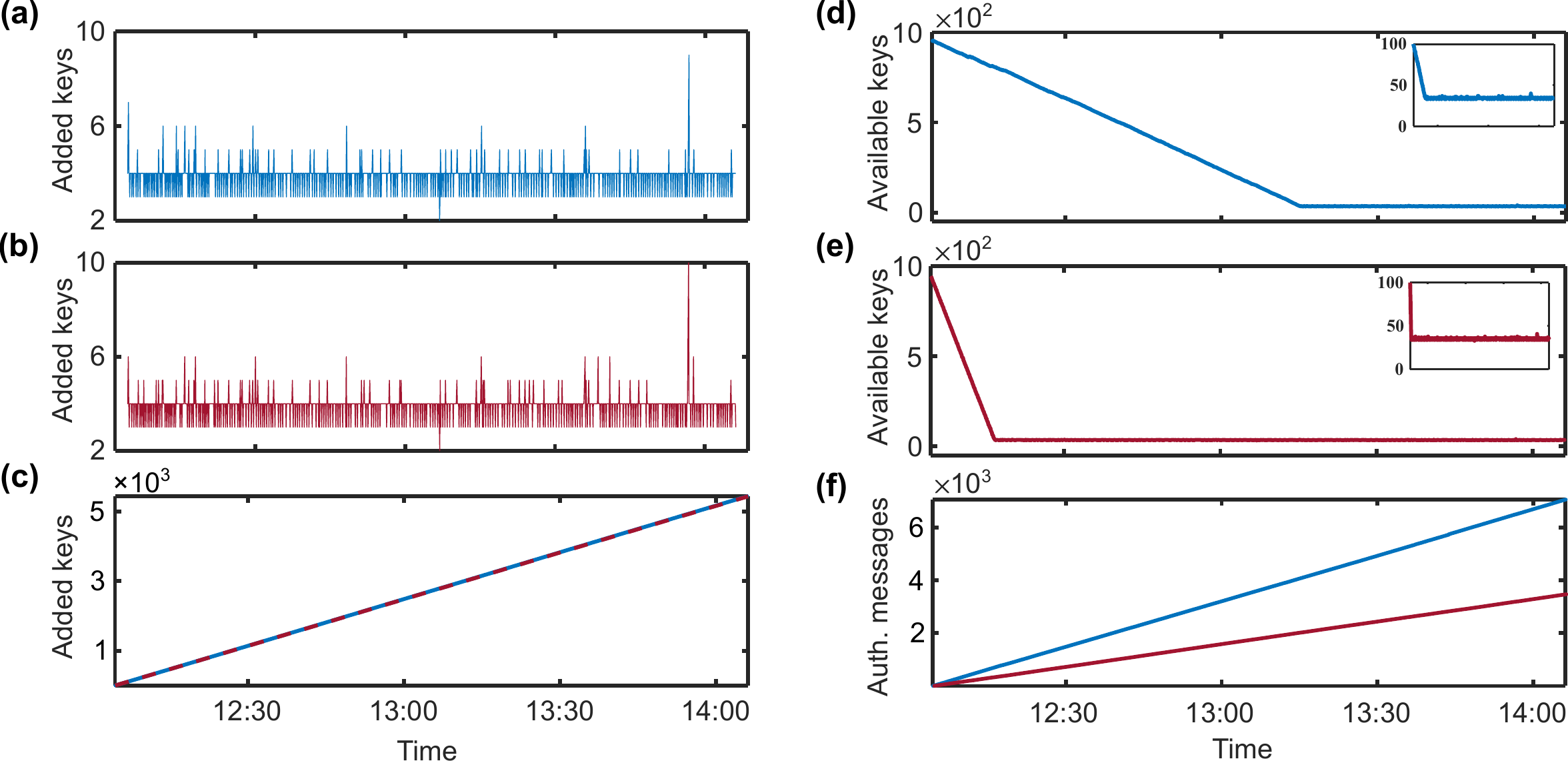}
\caption{Polled every 5 seconds 
(a)~Number keys added to the Intel Agent system. 
(b)~Number keys added to the Pv Agent system. 
(c)~Total number of the added keys to the Intel (blue) and Pv (red) agents.
The number of keys available to 
(d)~The Intel agent and 
(e)~The Pv agent. The inset figure shows the minimum keys that each agent must maintain (30 keys, which we chose arbitrarily) to synchronize the secret keys. 
(f)~The number of authenticated messages by the Intel (blue) and PV (red) agents.}
\label{results}
\end{figure*}

We collect data related to the number of added and available keys for each agent. Figure~\ref{results}(a) and Figure~\ref{results}(b) show the number of added keys for the Intel and the PV agent, respectively, as a function of time during the demonstration. The number of added keys is reported by the agent every 5 seconds---to conserve processing time required to check more frequently. If a more frequent key material update was used, both figures would be identical. The drop after 13:00 and spike before 14:00 in the added keys are likely the results of environmental changes, including wind gusts affecting the aerially deployed fiber. Figure~\ref{results}(c) shows the total number of keys added by each agent. Further, to avoid one node using a key that is not yet polled by the other node---due to polling delay, synchronization delay, or network disruption---we set a lower threshold $T$ on the number of keys each node keeps as a \textit{reserve pool}. In this work, we set $T=30$ as the minimum number of keys each node must keep before using a new key.  
Hence, we keep using the last known secret key---always with a new initialization vector---until the threshold $T>30$, then a new key is used. The key reuse typically lasts for approximately 5 seconds until the subsequent key poll is complete.
Figure~\ref{results}(d) and Figure~\ref{results}(e) show the number of available secret keys at each agent as a function of time. Before starting the energy storage system communications, each agent starts collecting keys from the QKD system. When the agents start communicating, a reservoir of approximately 950 keys is available in the secret key file. Then, each begins authenticating their received messages using an odd (or an even) key identification number for the Intel (PV) agent. Figure~\ref{results}(d) and Figure~\ref{results}(e) shows comparatively slower key consumption by the Intel agent compared to the PV agent. This slower consumption is due to their functional differences resulting in a difference in the rate of sent messages. Consequently, as shown in Figure~\ref{results}(f), the PV agent authenticates messages at a slower rate. 

This paper presents the first demonstration of quantum key-based authentication of smart grid communications across an energy delivery infrastructure environment. The developed system utilizes a flexible and scalable smart grid communications protocol: a publish-subscribe method. Further, keys from a commercial Qubitekk quantum key distribution system along with the Carter-Wegman authentication protocol are used, which in principle offer information-theoretic security. With this demonstration, quantum and classical security technologies have been shown to work in the energy infrastructure to authenticate data and control communications, providing long-term security, capable of exceeding the expected infrastructure service life. Future development of the reported techniques could include full hardware integration via smart grid manufacturers. In addition, hardware platforms with fully integrated power electronics systems are in development today in a new facility called the Grid Research Integration and Deployment Center (GridC). This facility provides an avenue to fully scale the presented implementation into multiple power electronics systems and integration demonstrations. On the other hand, in terms of cybersecurity, previous work demonstrated the trusted relay on the power grid but stopped short of showing how to use the distributed secret keys~\cite{Evans2021}, which is the focus of this work. Future work could concentrate on developing scalable secure communications including a wider range of power infrastructure devices.

\section*{Discussion}
Cyberattacks seeking to disrupt grid communications can have devastating consequences for grid operations. Therefore, verifying that the grid communications have originated from the authorized user is crucial. One way to authenticate information in transfer over a network is by employing an authenticator that can be used as a challenge to verify the authenticity of a message. Several methods are available to produce a message authenticator: message encryption, hash functions, or a message authentication code (MAC). Message encryption uses symmetric or asymmetric cryptographic algorithms. In light of the latency issues evident with public-key cryptography as outlined earlier, QKD secret keys for symmetric cryptography offer an attractive solution for long-term secure grid communications authentication.
On the other hand, message encryption hides information, and only users who know the secret key can encrypt and decrypt a message. Furthermore, for smart grid communications, the information in transit contains typical measurement data such as voltage, current, frequency, and phase that need to be examined for correctness---but not necessarily encrypted. As a result, in some applications such as the distribution automation system~\cite{Lim2010} authentication is preferable to encryption as data is usable during troubleshooting (such as a delay) with the cryptographic operations. Additionally, authentication has a further advantage in requiring fewer random bits from the QKD than full data encryption.

While it would be possible to deploy free-space terminals to perform QKD, the availability of fiber optic infrastructure makes for a much more convenient alternative, as one does not need to worry about objects (such as inclement weather) blocking the communications path. The presence of much higher power classical optical signals on the same fiber as the quantum signals can introduce a considerable amount of noise in the quantum channel~\cite{Peters2009, Mao2018}. As a result, it is highly advantageous to utilize a ``dark'' fiber dedicated to the quantum signal alone. Fortunately, many utilities---as part of the power grid modernization---have been heavily investing in information technology, including deploying optical fibers between operations centers, substations, and distributed energy resources. The investment in fiber offers the utility companies considerable bandwidth, which can be partially leased, as well as flexibility for grid operational communications~\cite{Elliott2002}.

The North American Electric Reliability Corporation (NERC) issued a set of Critical Infrastructure Protection (CIP) reliability standards~\cite{NERC} to ensure the security of Bulk Electric System (BES). The Physical Security Perimeter (PSP) standard (CIP-006-6) defines a physical security plan to safeguard BES cyber systems against any compromise that might cause improper BES behavior. For this reason, PSP access control requirements include key card access, special locks, security personnel, and authentication devices such as biometrics and tokens. Also, the standard outlines methods to monitor and log the physical access using alarm systems, human observation, computerized logging, and video surveillance and recording, which guarantees the physical security of the systems. However, the connectivity between a utility SCADA system and devices is the current widespread networking suite called Transmission Control Protocol/Internet Protocol (TCP/IP)~\cite{Cerf1974, braden1989}. While all communications for the electrical system must be trustworthy~\cite{Metke2010} and timely~\cite{Kansal2012}, transmitting data via a TCP/IP protocol is susceptible to cyber-attacks including spoofing~\cite{Wei2011}. Such attacks include injecting malicious data during transmission that may result in poor control responses and outages could occur. For this reason, electrical systems connected to the internet are potentially vulnerable to cyber-attacks~\cite{Wang2011, Kansal2012}.

Authentication of data and control messages is crucial for reliable, safe, and secure grid operations. Using an authentication protocol and secret keys known only to the sender and the receiver enable bi-directional message authentication. Moreover, an information-theoretic (meaning security is not based upon computing resource assumptions) authentication protocol based on private-key encryption comes without the latency penalty of public-key cryptosystems~\cite{Scarani2009, Barker2015}. For example, using the Carter-Wegman~\cite{Wegman1981} authentication protocol requires fewer computational resources and thus provides a long-lasting and more resource-efficient authentication compared to the asymmetric public-key-based authentication protocols~\cite{Hauser2012}. Thus, Demonstrating QKD technology in a real-world environment to verify the feasibility of quantum-based cybersecurity for power grid communications is a crucial way point towards wider adoption. A controlled laboratory setup dramatically reduces environmental impacts compared to field deployments. For example, environmental variables such as temperature and humidity, in addition to the electromagnetic emanations of specialized power equipment, can affect the quantum hardware, including optics, electronics, and electro-optics. Further, the fiber optic deployment mechanism in a real-world environment is another vital element to consider. The QKD key rate of an underground and aerial fiber will likely be affected in some QKD implementations and may require additional equipment/engineering compared to lab-based demonstrations.

MQTT~\cite{Mqtt-v5} is a communications protocol based on the publish-subscribe model (instead of the typical client-server architecture) developed in 1999 to minimize power and bandwidth requirements~\cite{Birman1987}. In the publish-subscribe communications paradigm, the publishers and subscribers never communicate directly but utilize a third-party intermediary, commonly referred to as a \textit{broker}. The broker's responsibility is to process all incoming traffic and appropriately deliver messages to the intended subscribers. As a result, this communications approach scales more effectively than the typical client-server architecture. An MQTT client can be a publisher or subscriber. The publisher role enables a client to send messages to the broker, who then relays them to the interested subscribers. Each published message must contain a required topic---that clients subscribe to its relevant messages---and an optional payload. For this reason, the broker activities can be parallelized---using topic-based filtering---in an event-based manner, making it an ideal protocol for IoT services.

\section*{Acknowledgements}
We acknowledge the support of Steve Morrison, Tyler Morgan, and Ken Jones and Patrick Swingle. This work was partially performed at Oak Ridge National Laboratory (ORNL). ORNL is managed by UT-Battelle, LLC, under Contract No. DE-AC05-00OR22725 for the DOE. We acknowledge support from the DOE Office of Cybersecurity Energy Security and Emergency Response (CESER) through the Cybersecurity for Energy Delivery Systems (CEDS) program.


%

\newpage

\onecolumngrid
\section*{Supplementary Information}

\begin{algorithm}[h!]
  \caption{Create MAC for each outgoing message}
  \label{algorithm1}
\raggedright\hspace*{\algorithmicindent} \textbf{Input} message, topic \\
\raggedright\hspace*{\algorithmicindent} \textbf{Output} payload
  \begin{algorithmic}[1]
    \Function{publish}{message}
    \State $m\gets message$ 
    \State $t\gets topic$ 
    \State $n\gets number~of~next~secret~key$ 
    \State $ts\gets timestamp$ 
    \State $tm\gets m+t+n+ts$ 
    \State $key\gets the~nth~secret~key~from~key~table$ 
    \State $iv\gets next~random~number~from~QRNG$ 
    \State $mac_S\gets GMAC_E(tm,key,iv)$ 
    \State $p\gets tm+iv+mac_S$ 
    \State \textbf{return} $p$ 
    \EndFunction
  \end{algorithmic}
\end{algorithm}

\begin{algorithm}[h!]
  \caption{Verify MAC for each incoming message}
  \label{algorithm2}
  \raggedright\hspace*{\algorithmicindent} \textbf{Input} payload, topic \\
  \raggedright\hspace*{\algorithmicindent} \textbf{Output} True/False
  \begin{algorithmic}[1]
    \Function{on\_message}{payload, topic}
    \For{$p$ in $payload$}
    \State $tm\gets p[1]$ 
    \State $iv\gets p[2]$
    \State $mac_s\gets p[3]$ 
    \EndFor
    \For{$q$ in $tm$}
    \State $m\gets q[1]$ 
    \State $t\gets q[2]$
    \State $n\gets q[3]$
    \State $ts\gets q[4]$
    \EndFor
    \State $ct\gets current~time$ 
    \State $key\gets the~nth~secret~key~from~key~table$ 
    \State $mac_R\gets GMAC_D(tm,key,iv)$ 
    \If{$mac_S = mac_R~\&~t = topic~\&~ts < ct - \delta$} 
    \State $result\gets True$ 
    \Else 
    \State $result\gets False$ 
    \EndIf
    \State \textbf{return} $result$     
    \EndFunction
  \end{algorithmic}
\end{algorithm}

\end{document}